# Magnetic properties of undoped and 15%F doped SmFeAsO compounds

M.R. Cimberle[1], C. Ferdeghini[2], F.Canepa[1,3], M.Ferretti[2,3], A. Martinelli[2], A.Palenzona[2,3], A.S. Siri[2,4], M. Tropeano[2,4]

[1]*CNR-IMEM via Dodecaneso 33, 16146 Genova – Italy*

[2]*CNR-INFM-LAMIA Corso Perrone 24, 16152 Genova – Italy*

[3]*Dipartimento di Chimica e Chimica Industriale, Università di Genova via Dodecaneso 31, 16146 Genova – Italy*

[4]*Dipartimento di Fisica, Università di Genova via Dodecaneso 33, 16146 Genova – Italy*

## Abstract

In this paper the magnetic behaviour of SmFeAsO and SmFeAs($O_{0.85}F_{0.15}$) samples is presented and discussed.

Molar susceptibility of SmFeAsO exhibits a local peak at T~140K due to the establishment of a long range antiferromagnetic ordering of the Fe moments in Fe-As layers. This feature has already been observed with different techniques, and frequently ascribed to the onset of a Spin Density Wave (SDW). At $T_N \cong 6K$ another peak, which we attribute to the establishment of antiferromagnetic ordering of Sm ion sublattice, is observed. Furthermore, a temperature independent signal (Pauli paramagnetism, Landau and core diamagnetism….) is also present in the magnetic behaviour of this sample. In SmFeAs($O_{0.85}F_{0.15}$) the antiferromagnetic ordering in Fe-As plane is suppressed and superconductivity occurs at T = 52 K, whereas the antiferromagnetic ordering of Sm ions at low temperature persists, leading to the coexistence and competition between superconducting and magnetic orderings. Above the transition temperature, after the subtraction of the Sm ion sublattice paramagnetic contribution and of the temperature independent contribution to the experimental susceptibility data, a Curie-Weiss behaviour for Fe is observed, with a magnetic moment of $1.4\mu_B$.

## I: Introduction

The recent discovery of superconductivity at temperatures up to T=29K in the iron oxipnictide LaFeAs($O_{1-x}F_x$)[1] has stimulated a lot of work on these materials. In short period of time different approaches have been attempted to increase the transition temperature. On the one hand La was substituted with other Rare Earth (RE) with smaller ionic radius (like Pr, Nd, Sm, Gd) in order to induce chemical pressure[2,3,4,5], on the other hand the optimal doping for the superconducting phase was studied and obtained with a variety of different techniques, i.e. F substitution on O site[1,6], O deficiency[7,8,9], and partial substitution on the RE site with bi-and tetravalent cationic species[10,11,12]. By combining these two approaches, good results have been achieved, so that $T_c$ has been increased up to the considerable value of about 55K[2,3,4,7,9]. Simultaneously, many different structural and physical characterization have been performed, together with a remarkable theoretical effort devoted to the comprehension of the mechanism of superconductivity and of the possible coexistence of superconductivity and magnetism. In the parent compound a structural distortion, , which usually precedes a long range antiferromagnetic ordering, is observed at T~150K[6,13,14]. Such antiferromagnetism has frequently been related to the onset of a SDW, also because of the particular behaviour of the related physical properties: resistivity, specific heat, Hall effect and far infrared reflectance, all suggesting the opening of an energy gap[15,16].

One of the most intriguing features of this class of materials is related to their magnetic properties. Similarly to what happens in high $T_c$ superconductors, the parent compound of these materials is anti-ferromagnetic; after doping, the anti-ferromagnetic ordering is suppressed and the compound exhibits a good metallic behaviour down to the superconducting transition temperature. Whether magnetic fluctuations are involved in the development of superconductivity is an open question that could give new insight on the analogous problem related to high $T_c$ superconductors. Moreover, what type of magnetism is present in the doped samples, both in the normal and superconducting state, is the object of a lively debate [17,18,19]. Furthermore, when La is substituted with other RE (Pr, Nd, Sm, Gd) another source of magnetism is introduced, although in the charge reservoir layers, not in the superconducting ones. Magnetic measurements are commonly used to characterize the superconducting behaviour of this new material, but few works have been devoted to study the normal state properties both in parent and doped samples. Uemura and coworkers[6] in LaFeAs($O_{1-x}F_x$) observed a Curie-Weiss behaviour above $T_c$ and a strong variation of the overall magnetic moment for different x values, with a maximum for x=0.05. Wang and coworkers[10] in GdOFeAs observed the magnetic moment dominated by the paramagnetic behaviour of Gd ions that order antiferromagnetically below 4.2 K. Tarantini et al.[20] measured NdFeAs($O_{0.89}F_{0.11}$) up to a magnetic field of 33 Tesla and argued that the magnetic behaviour is determined mostly by the Nd ions and, more important, that superconductivity and antiferromagnetism cannot coexist.

With the aim of giving a contribution to the outlined questions, DC susceptibility measurements, carried out on both SmOFeAs and SmFeAs($O_{0.85}F_{0.15}$) samples, are hereafter presented.

**II: Samples preparation and characterization**

SmFeAsO and SmFeAs($O_{0.85}F_{0.15}$) were both prepared in two steps: 1) synthesis of SmAs starting from pure elements in an evacuated glass flask at a maximum temperature of 550°C; and 2) synthesis of the oxypnictide in an evacuated quartz flask reacting SmAs with stoichiometric amounts of Fe, $Fe_2O_3$ and $FeF_2$ (details concerning the parent sample are reported in[21]). The second step for SmFeAsO and SmFeAs($O_{0.85}F_{0.15}$) was carried out at 1200°C for 24h and 1000°C for 15h, respectively.

The samples were characterized by X-ray powder diffraction and by scanning electron microscopy (SEM)[21]. The XRPD pattern confirms the formation of both SmFeAsO and SmFeAs($O_{0.85}F_{0.15}$); $Sm_2O_3$ can hardly be detected as impurity in the parent sample, whereas few amounts SmOF are present in the F-substituted one (in any case less than 2%). After F-substitution the cell size contracts, with the decrease along the c-axis more enhanced. SEM observation reveals that both samples are constituted of connected micrometric crystals: in SmFeAsO rectangular shaped tabular crystals of with edge up to 20 μm are observed, whereas in SmFeAs($O_{0.85}F_{0.15}$) the tabular crystals are characterized by a polygonal edge up to 5 μm. It is not clear if these differences in the crystalline shape are dominated by F-substitution or the different reaction temperature.

The resistivities of the two samples, normalized to the values at 300K, are reported in Fig. 1. The parent sample exhibits a pronounced anomaly around T~140 K, then resistivity decreases monotonically with decreasing temperature. Saturation is not achieved even at the lowest temperature (400mK): below about 6 K a drop is clearly visible.

The doped sample is characterized by a nearly linear decrease of resistivity with temperature, with $T_c$ =53.5 K, if taken as the onset of the derivative variation, and $T_c$ =51.5 if estimated from the maximum of dρ/dT. This temperature roughly corresponds to the critical temperature measured magnetically at low field ($T_c$=52.0 K at 10 Oe).

All the magnetic measurements were performed by a SQUID magnetometer ( MPMS by Quantum Design).

### III: Magnetic measurements
#### A. SmFeAsO

The molar susceptibility of SmFeAsO measured from 2K up to 300K is shown in Fig.2a. In this measurement, after a Zero Field Cooling (ZFC) procedure, a field of 30kOe was applied. The main features in Fig.2 are the following: i) at T~140K a local maximum is present, related to the antiferromagnetic ordering of Fe moments, ii) at T=6K a very sharp cusp is seen, showing the antiferromagnetic ordering of the Sm sublattice, in agreement with the sharp peak in the specific heat observed by Ding et al at 4.6K[22] and by Tropeano et al.[23] in the same sample presented here, and iii) a temperature dependent susceptibility signal superimposed on a **great** temperature-independent background one is present: the former varies from $\chi=4.95 \cdot 10^{-3}$ emu/mol at T= 6K up to $\chi=3.7 \cdot 10^{-3}$ emu/mol at T=300K, exhibiting a Curie-Weiss type behaviour. We argue that part of the magnetic background is due to ferromagnetic impurities that are present in our sample, although in very **small** amounts. This can be seen from magnetization versus field measurements at T=5K and T=300K and up to 50 kOe, displayed in the inset of Fig.2a. Here, a signal that linearly increases with the field is superimposed to a ferromagnetic signal, sharply increasing at low field and saturating around 10 kOe to about the same value ( the extrapolated magnetization values are 80 and 70 emu/mol at T= 5K and 300K respectively), indicating that the ferromagnetic impurities have a transition temperature higher than room temperature. This ferromagnetic contribution has been subtracted to the measured curves, yielding the data of molar susceptibility shown in Fig.2.b. The overall magnetic behaviour is in very good agreement with the resistivity curve of SmFeAsO in Fig.1, where drops are observed at 140K and 6K.

The continuous line in Fig. 2b is a fit of the experimental curve obtained by taking the different contributions to the magnetic moment into account: 1) the paramagnetic contribution ($\chi_{Sm}$) of Sm ion lattice, 2) the magnetic contribution ($\chi_{Fe}$) due to Fe sublattice in the Fe-As layers which is antiferromagnetic below T=140K and paramagnetic above, and 3) a temperature independent contribution ($\chi_0$) due to the Pauli paramagnetism and to any diamagnetic signal. Therefore:

$$\chi = \chi_0 + \chi_{Sm} + \chi_{Fe} \tag{1}$$

Let us analyse each component of $\chi$ in detail.

1) $\chi_{Sm}$: The RE Sm presents in most of its compounds an anomalous susceptibility behaviour, displaying a more or less pronounced minimum at temperatures around or above room temperature: consequently the Curie-Weiss behaviour is not observed. This trend has been successfully ascribed[24] to the fact that in the $Sm^{3+}$ ion the J multiplet intervals are comparable with $kT$: so, not only the J = 5/2 ground state is occupied but also the first exited level J=7/2. As a consequence, the sum of the J levels, stopped to the second term and taking the Boltzmann temperature factor into account, gives :

$$\chi_{Sm} = x \cdot \frac{C1}{(T-\vartheta_{Sm})} + (1-x)\frac{C2}{(T-\vartheta_{Sm})} \cdot e^{(-\frac{\Delta}{T})} \tag{2}$$

where x represents the fractional occupation of the two states, θ is the paramagnetic Curie temperature, C1 =0.0903 emu K/mol and C2=1.3452 emu K/mol are the calculated Curie constant related to the two J states, and $\Delta = E_j/k$ is the difference in temperature between the two states.

2) $\chi_{Fe}$: Above the antiferromagnetic transition temperature the contribution of the iron sublattice is paramagnetic and we can write:

$$\chi_{Fe} = \frac{C3}{(T - \vartheta_{Fe})} \qquad (3)$$

where C3 is the Curie constant related to the magnetic moment carried by the $Fe^{2+}$ ion, and $\theta_{Fe}$ is the paramagnetic Curie temperature. Below the antiferromagnetic transition temperature, the magnetic susceptibility of the Fe sublattice was treated in the framework of the Mean Field Theory (MFT) for antiferromagnetic systems[25]. Here, in a polycrystal compound the magnetic susceptibility is due to the two contributions:

$$\chi_{poly.} = \tfrac{2}{3}\chi_\perp + \tfrac{1}{3}\chi_\parallel \qquad (4)$$

where $\chi_\perp$ is the component with the magnetic field applied perpendicular to the magnetic sublattice, and $\chi_\parallel$ is the contribution obtained when the applied field is applied parallel to the sublattice. In the MFT, as a first approximation, the $\chi_\perp$ contribution is constant (and equal to $\chi(T_N)$ for $T_N > T > 0K$) while the $\chi_\parallel$ temperature dependence, in the same T span, has a $T^2$ behaviour. Therefore for $T<T_N$ we used

$$\chi_{Fe} = 2/3\chi_{Fe}(T_N) + A \cdot T^2$$

(5)

3) $\chi_0$: The $\chi_0$ term covers all the different temperature independent contributions (Pauli paramagnetism and Landau diamagnetism of the conduction electrons, high frequency contribution[23], diamagnetic contributions arising from nuclei and electronic inner shells ...).

The obtained fit is satisfactory: it reproduces the main aspects of the experimental data with reasonable values of the parameters that are reported in table I. Small differences in the details of the fit may be ascribed to the presence of very small quantity of impurities, mainly the paramagnetic SmOF, which is less than 2%, and, in even smaller amount, the antiferromagnetic $Fe_2As$ ($T_N$=353K) and FeAs ($T_N$=77K). We point out that the C3 value (0.025 emu K/mol) corresponds to a magnetic moment of $0.4\mu_B$, compatible with the values reported in the literature, ranging from $0,25\mu_B$ up to $0.35\mu_B$[13,26]. Moreover, the relatively high positive value of the temperature independent term should be emphasized here ($\chi_0$= 0.94 $\cdot 10^{-3}$ emu/mol). This value is

mainly ascribed to the Pauli positive contribution arising from non interacting band electrons, written as:

$$\chi_P = 2N(E_F)\mu_B^2 \quad (6)$$

where $N(E_F)$, the density of states at the Fermi level, is proportional to the effective mass $m^*$ of the electrons through

$$N(E_F) = \frac{(2m^*)^{3/2} E_F^{1/2}}{4\pi^2} \quad (7)$$

For normal metals, such as Cu, $\chi_0$ is around $10^{-5} \div 10^{-6}$ emu/mol, which corresponds to an effective mass m* comparable with the mass of the free electron. In this case $\chi_0$ (~$10^{-3}$ emu/mol) is about one hundred times the Pauli susceptibility of typical metals, such as Cu. So, the possibility that 4f electrons of Sm can hybridize with the conduction electrons, giving a strong enhancement of carrier effective mass (*heavy fermion state*) can be taken into account. Values of $\chi_0$ as high as this one or more were found in typical heavy fermion or Concentrated Kondo systems ($\chi_0$ ($CeCu_2Si_2$) = 6.5 ·$10^{-3}$ emu/mol[27], $\chi_0$ ($CeAl_3$) = 3.6·$10^{-3}$ emu/mol[28]). Recent measurements of heat capacity in SmOFeAs gave evidence of a high value of the electronic coefficient γ of the specific heat ( the measured γ value varies from 42[23] to 119.4[22] mJ/K$^2$ mol). These high γ values are consistent with our high $\chi_0$ value confirming, thus, the possible *heavy-fermion* character of the system.

## B. SmFeAs($O_{0.85}F_{0.15}$)

### B.I. Normal State

The molar susceptibility of the SmFeAs($O_{0.85}F_{0.15}$) sample from T=2K up to T= 300K is shown in Fig.3. Also in this case a magnetic field of 30kOe for ZFC and FC procedure was applied. The main features that may be observed in this figure are the following: i) at $T_c$ ~ 50K a very sharp decrease of magnetization marks the superconducting transition, ii) above $T_c$ a monotonous decrease of magnetization for increasing temperature is observed up to 300K and no local maximum is present around T~140K, and iii) at low temperature a complex behaviour in ZFC-FC measurements on the superconducting side may be observed, which we will discuss in the next paragraph. In the inset the magnetization measured at T=60 K and T=300K is shown. Unlike what is observed in the parent sample, at both temperatures we see only a signal that increases linearly with the field. The low field extrapolation is zero, therefore no procedure of subtraction of a ferromagnetic background signal was required, which indicates a good level of purity in this sample. As reported in "Sample preparation and characterization", in this sample the main impurity phase detected with SEM analysis is SmOF (less than 2 %). A sample of SmOF was therefore prepared by mixing equal quantities of $Sm_2O_3$ and $SmF_3$ in a sealed quartz ampoule and reacting at 850°C. Its magnetic susceptibility was measured in the same temperature and field range as $SmFeAs(O_{0.85}F_{0.15})$. SmOF is paramagnetic in the whole temperature range and above 50 K its contribution is less than 1/100 of the overall magnetic susceptibility, so we neglect it in the analysis of the magnetic data.

If we compare this measurement with that shown in Fig.2b for the parent compound we note two important differences: the local maximum at T~140K, indicating the antiferromagnetic ordering is lacking, and the overall magnetic signal greatly increased (for example here, at T= 60K, $\chi= 3.4 \cdot 10^{-3}$ emu/mol versus $\chi= 1.57 \cdot 10^{-3}$ emu/mol for the parent sample). The vanishing of the high temperature magnetic ordering and a comparable increase in the Fe magnetic moment had already been observed by Nomura et al.[6] in their analysis on F substitution effects on the crystallographic and magnetic properties of $LaFeAs(O_{1-x}F_x)$. In the same Fig.3 the continuous line represents a fit of $\chi$ that has been done using the relationship 1) $\chi = \chi_0 + \chi_{Sm} + \chi_{Fe}$ where $\chi_0$ and $\chi_{Sm}$ are the same used in the fit for the parent sample and $\chi_{Fe}$ has now a Curie-Weiss paramagnetic behaviour in the whole temperature range above $T_c$. The fit is good and the result obtained is $\chi_{Fe} = \frac{C_3}{(T-\vartheta)}$, where $C_3$= 0.2644 emu K/mol corresponding to a magnetic moment $\mu_{Fe}$=1.4 $\mu_B$, and $\theta_{Fe}$= -82.6K. All the fit parameters are shown in Table I. The partial substitution of O by F has different effects. The first is an increase in the electron concentration: as observed by Si et Abrahams[29], there is a change in the spin state of Fe because the Fe electron goes into one of the d-bands existing in the Fe-As layers. The second effect is the reduction of the lattice parameter mainly along the c-axis, which is related to a change in the inter-layer bonding.

Thus a strong perturbation of the spin density distribution in the system is possible, causing both the destruction of the antiferromagnetic ordering and the increase in the Fe magnetic moment. Examining Table 1 we note that, regarding the Sm ions, a small difference is seen in the Δ and x values relative to $\chi_{Sm}$. The Δ variation may be related to the c axis decrease and, as a consequence of the change of energy span between the two levels, also the relative population of them is slightly changed.

### B.II. Superconducting State

In Fig.4 an enlargement of Fig. 3 in the low temperature region is shown to observe the superconducting behaviour in detail: the four curves correspond to ZFC-FC measurements made starting from T=2K (open symbols) or from T=5K (filled symbols). At T~50K the magnetization begins to decrease, due to the diamagnetic signal of the superconducting state, but, as temperature is lowered, a minimum occcurs at about T=25K, and then the signal increases. The ZFC and FC measurements are coincident from $T_c$ down to about the temperature of the minimum (~25 K), which indicates a large zone of reversibility in the H-T phase diagram. At lower temperature the curves separate **from each other**: the FC curves continue to increase, but a slightly higher value is observed for the measurement that started from 5K, while the other one increases down to 4K, then shows a slightly decreasing behaviour down to the minimum temperature of 2K. The two ZFC curves show the expected shielding but also, in this case, different behaviours that depend on the measurement starting temperature. This anomalous behaviour, and in particular the difference between the two ZFC curves, is not fortuitous: we have repeated the measurement many times, changing the applied magnetic field and on different samples, but the result does not change. It may be understood having in mind that Sm ion sublattice orders antiferromagnetically at low temperature, as we have observed in the parent sample (see Fig.2). Such ordering has been evidenced by Ding and co-workers[22] in both SmFeAsOF and $SmFeAsO_{0.85}F_{0.15}$ samples by specific heat measurements. In particular they observe a peak at T= 4.6 K in the SmFeAsO sample, while in the superconducting one the peak is shifted to T= 3.7 K. Below this temperature superconductivity and RE lattice antiferromagnetism coexist. The same Neel temperatures were found by specific heat measurements by Tropeano et al.[23] on the same parent sample here presented and a 7%F doped one. The ZFC curve initiated at 5K might be seen simply as sum of diamagnetic signal of the superconductor and magnetic signal of Sm lattice. The ZFC curve initiated at T=2K indicates that, starting from a temperature just below $T_N$, the shielding signal abruptly changes passing at $T_N$ through the maximum of magnetization of the Sm sublattice, and gives rise to a sort of sharp corner, before recovering the behaviour of the other ZFC curve. It is not simply a matter of summing the two signals, because the second measurement indicates that one signal (the magnetic one) changes the superconducting one. Similarly, in FC measurements the difference may be ascribed to the different internal field felt by the superconductor if the measurement starts from $T<T_N$. Finally, the slightly increasing FC signal observed from 2K up to 5K depends on the competition between a FC signal at variable internal field and the magnetic signal of Sm lattice.

Very recently a study of the magnetic properties of $NdFeAs(O_{0.89}F_{0.11})$ sample has brought the authors [20] to the conviction that the magnetic behaviour is dominated by Nd ions and that no long-range antiferromagnetic ordering of Nd sublattice exists below $T_c$, In fact magnetization versus temperature measurements performed from T=4.2 up to 20K always exhibit paramagnetic behaviour. We suggest that Nd ion lattice orders antiferromagnetically at temperature lower that 4.2 K and much lower than the Curie temperature of about 10-11K, which the authors estimated from the Curie-Weiss behaviour at high temperature. In fact, very recent measurements by Y.Qiu et al.[30] observed AF ordering below 2K in NdFeAs($O_{0.8}F_{0.2}$). Sm ions sublattice orders at temperature higher than Nd (in agreement with the De Gennes factor) and RE antiferromagnetism and superconductivity coexist in this new class of superconducting compounds as in many other classes of superconducting and magnetic materials (RERh$_4$B$_4$, borocarbides, ruthenocuprates, ….). We recall that superconductivity and RE magnetism are located in different planes of the unit cell in SmFeAs($O_{1-x}F_x$). On the other hand, the problem of the magnetism in the FeAs planes, where superconductivity sets in, is a very intriguing and totally open question.

### IV: Conclusion

The magnetic characterization of the parent SmOFeAs sample and 15% doped SmFeAs($O_{0.85}F_{0.15}$) sample has been performed and discussed.

In the parent sample, in addition to the well known antiferromagnetic ordering of Fe sublattice, we observed the antiferromagnetic ordering of the Sm ions at T=6K. The magnetic signal has been fitted as the sum of the Sm and Fe sublattices, plus a temperature independent magnetic contribution mainly related to an enhanced Pauli paramagnetism. The Sm paramagnetic contribution was described in terms of different population of the ground and first excited level. In the parent sample a Fe magnetic moment of 0.4 $\mu_B$, comparable with values obtained with different techniques, was found. In the doped sample the Fe contribution in the normal state, obtained after subtracting from the measured signal the Sm ion magnetic signal and the temperature independent one, turns out to be a good Curie-Weiss behaviour with a magnetic moment of 1.4$\mu_B$.

The temperature independent term, $\chi_0$, observed in both samples, exhibits a relatively high value: about one hundred times the Pauli susceptibility of typical metals, such as Cu. We suggest that this high value is due to an enhancement of the carrier effective mass, as a consequence of the possible hybridization of 4f states of RE with conduction electrons, and indicates a possible heavy fermion character of the system.

Finally, at low temperature the antiferromagnetic ordering of Sm ions is seen through its effect on the coexisting superconducting state.

### Acknowledgments

We thank Compagnia di S.Paolo for the financial support.

**Figure captions**

**Fig.1** Resistivity versus temperature for SmFeAsO and SmFeAs(O$_{0.85}$F$_{0.15}$) samples. Data are normalized to the $\rho_{300}$ value.

**Fig.2** Panel a: molar susceptibility versus temperature for the SmFeAsO sample; the applied field was 30 kOe. In the inset: magnetization versus magnetic field measurements performed at T=5K and T=300K. Panel b: molar susceptibility versus temperature for the SmFeAsO sample where a ferromagnetic contribution valued by M versus magnetic field measurements has been subtracted. The continuous line is a best fit of the data by the relationships 1), 2), 3) and 5) (see text for details)

**Fig.3** molar susceptibility versus temperature for the SmFeAs(O$_{0.85}$F$_{0.15}$) sample. The applied field was 30 kOe. ZFC and FC measurements have been performed starting from T=2K (open symbols) or T=5K (filled symbols). In the inset magnetization versus magnetic field measurements performed at T=5K and T=300K are shown. The continuous line is a best fit of the data (see text for details).

**Fig.4** Enlargement of Fig. 3 in the low temperature region. Open symbols: ZFC-FC measurements performed starting from 2K; filled symbols: ZFC-FC measurements performed starting from 5K.

**Table caption**

**Table I.** Values of the magnetic parameters obtained from the molar susceptibility fits of the parent and doped samples.

**Table 1**

| Compound | $\chi_0$ (emu/mol) | x | $\Delta$(K) | $\theta_{Sm}$(K) | $\theta_{Fe}$(K) | $\mu_{Fe}$ ($\mu_B$) | A (emu/mol K$^2$) |
|---|---|---|---|---|---|---|---|
| SmFeAsO | 0.94×10$^{-3}$ | 0.7 | 520 | -49 | -30 | 0.4 | 8.0×10$^{-10}$ |
| SmFeAs(O$_{0.85}$F$_{0.15}$) | 0.94×10$^{-3}$ | 0.75 | 620 | -50 | -82.6 | 1.4 | – |

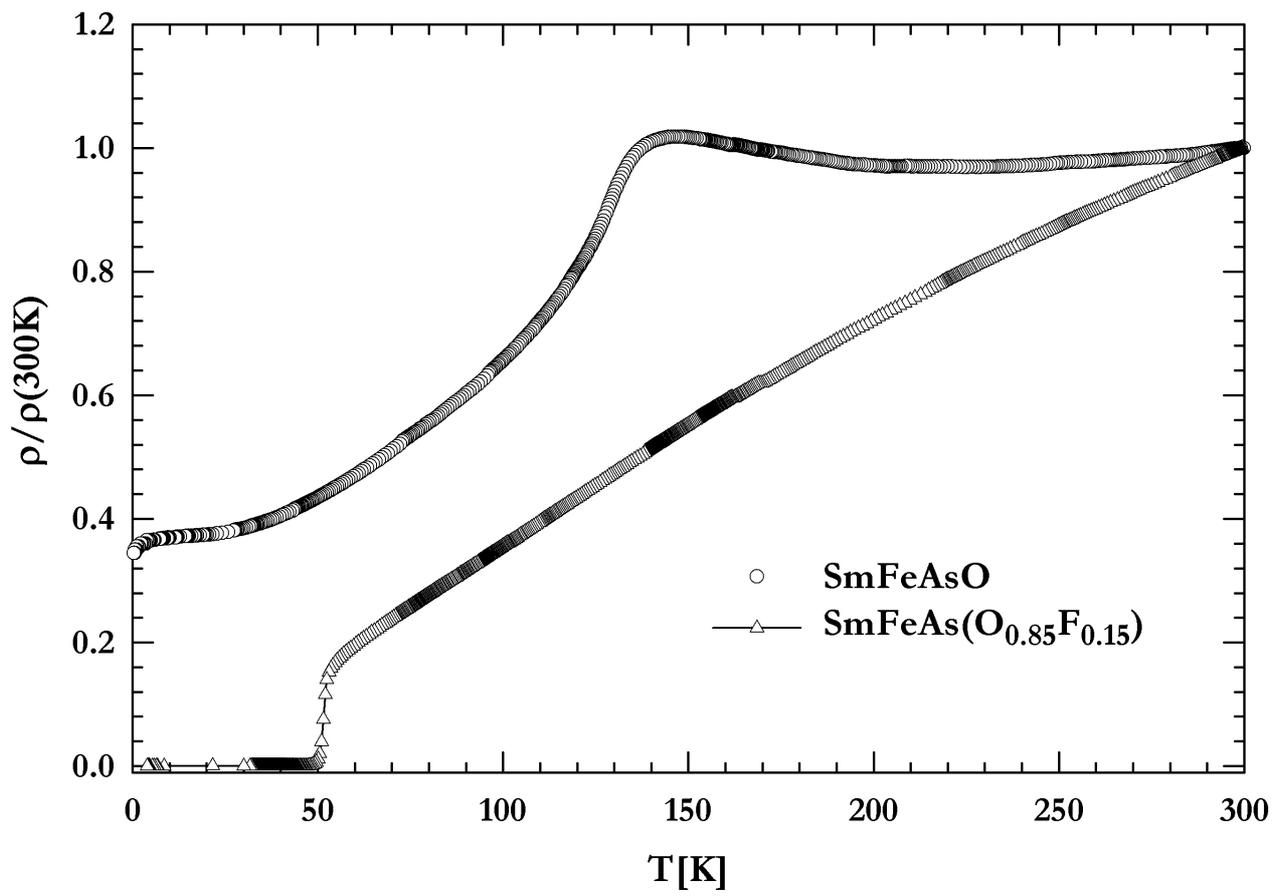

Fig.1

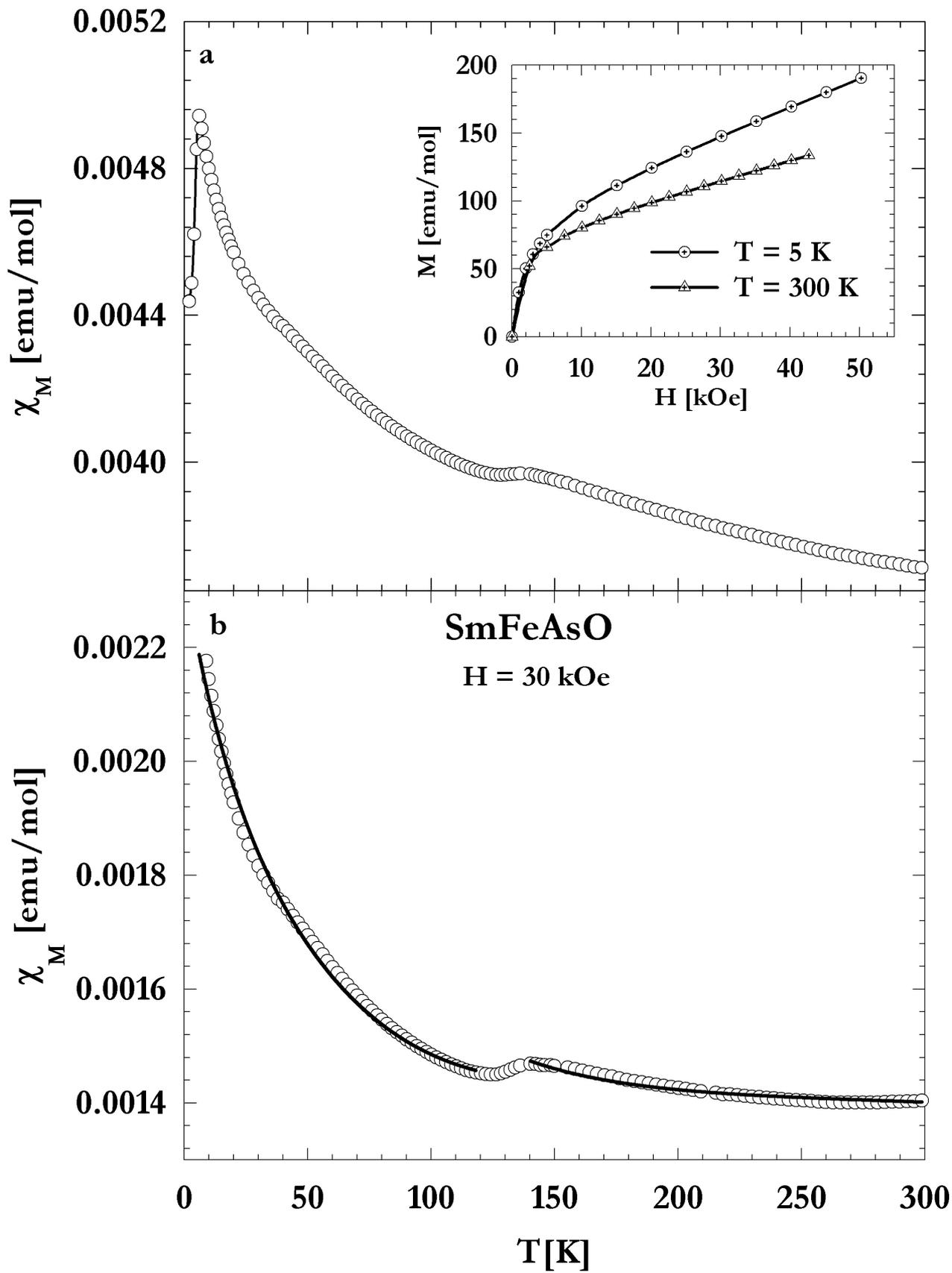

Fig.2

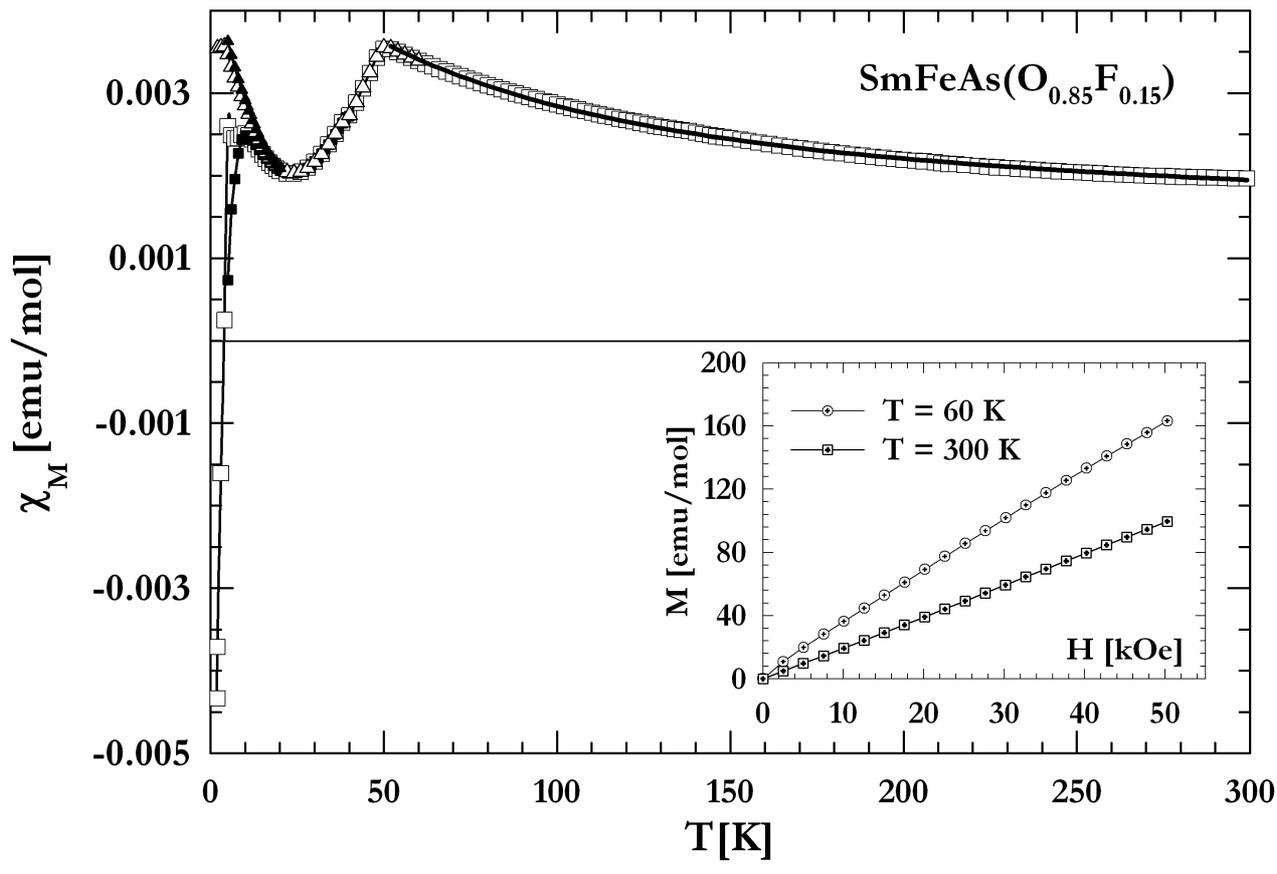

Fig.3

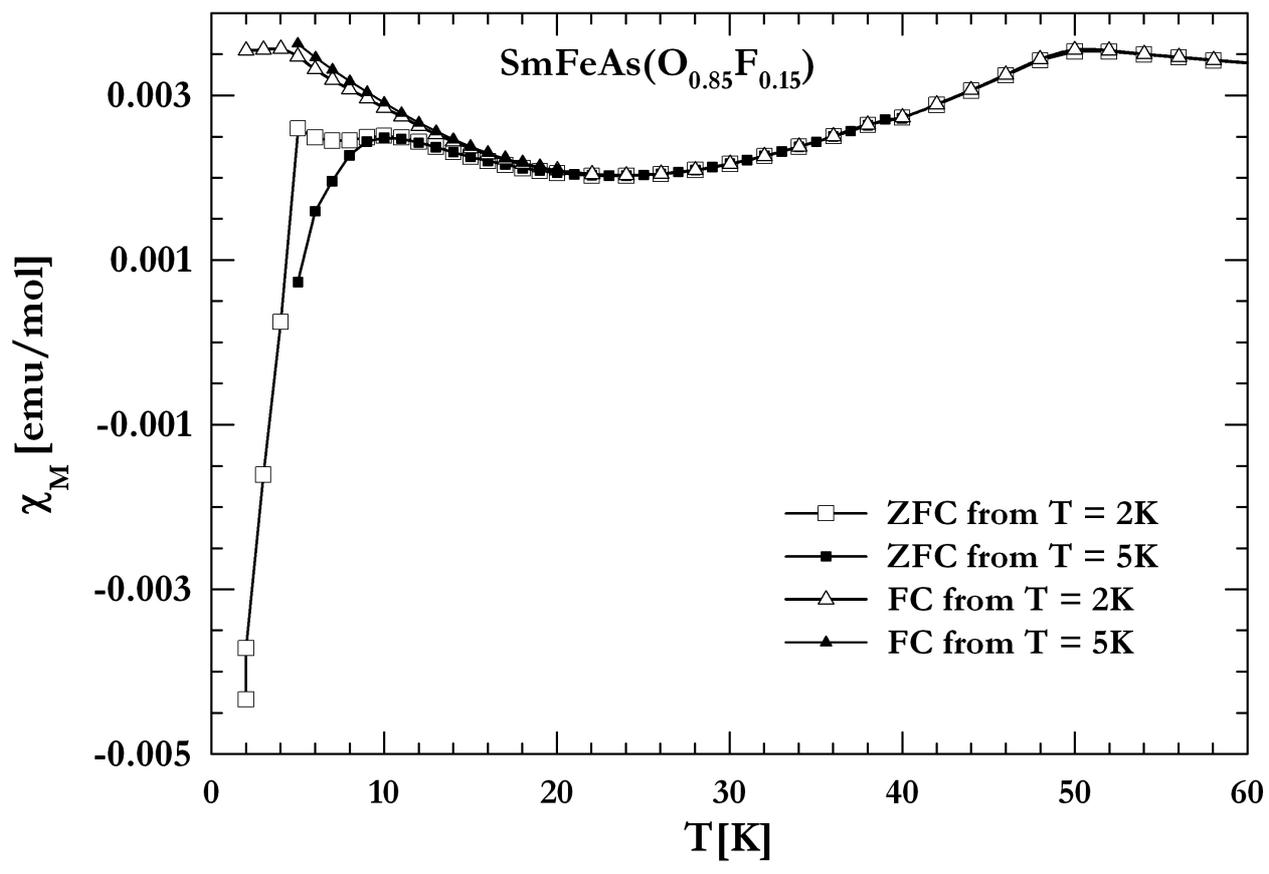

Fig.4